\documentclass[aps,a4paper,10pt,twocolumn,nofootinbib,floatfix]{revtex4} 
\usepackage{amsfonts}
\usepackage{amssymb}
\usepackage{mathrsfs}
\usepackage[mathscr]{euscript}
\usepackage[dvips]{graphicx}
\usepackage{fancyhdr}
\usepackage{makeidx}

\begin{document}

\title{Limits of the uni-directional pulse propagation approximation}

\author{P. Kinsler}
\email{Dr.Paul.Kinsler@physics.org}
\affiliation{
  Blackett Laboratory, Imperial College London,
  Prince Consort Road,
  London SW7 2AZ,
  United Kingdom.}

\begin{abstract}

I apply the method of characteristics to both 
 bi-directional 
 and uni-directional pulse propagation 
 in dispersionless media containing nonlinearity of arbitrary order.
The differing analytic predictions for the shocking distance
 quantify the effects of the 
 uni-directional approximation
 used in many pulse propagation models.
Results from numerical simulations support
 the theoretical predictions, 
 and reveal the nature of the coupling between 
 forward and backward waves.

\end{abstract}



\lhead{\includegraphics[height=5mm,angle=0]{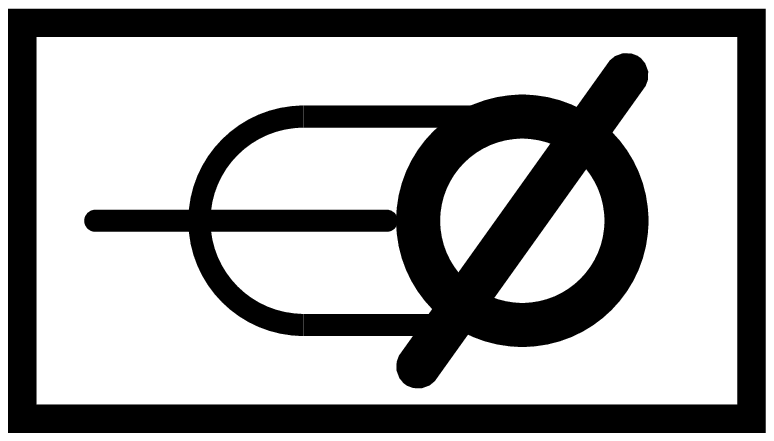}~~MOCNL}
\chead{~}
\rhead{
\href{mailto:Dr.Paul.Kinsler@physics.org}{Dr.Paul.Kinsler@physics.org}
}
\lfoot{ }
\rfoot{Kinsler-2007arXiv-mocnl}

\date{\today}
\maketitle
\thispagestyle{fancy}




{\em
\noindent 
  Published as J. Opt. Soc. Am. B {\bf 24}, 2363 (2007).
This version contains an additional appendix containing the 
MOC predictions for a time-propagated theory.
}

\section{Introduction}

Most approaches to optical pulse propagation rely on an 
 approximation where the fields
 only propagate forwards.
Even the recently derived extensions of 
 typical propagation methods used in nonlinear optics
 (e.g. 
 \cite{Brabec-K-1997prl,Kinsler-N-2003pra})
 assume a complete decoupling between 
 oppositely propagating fields
 to optimise the calculation.
Moreover,
 those based directly on Maxwell's equations
 (e.g. \cite{Kolesik-MM-2002prl,Tyrrell-KN-2005jmo,Kinsler-RN-2005pra})
 or the second order wave equation 
 (e.g. \cite{Blow-W-1989jqe,Ferrando-ZCBM-2005pre,Genty-KKD-2007ox}),
 are often simplified to work in the forward-only limit, 
 where backward propagating fields are set to zero.
This is despite 
 directional decompositions of Maxwell's equations 
 (e.g. \cite{Kinsler-RN-2005pra,Kinsler-2006arXiv-fleck})
 indicating that nonlinearity inevitably
 couples the forward and backward waves together --
 and even creates a backward field if one is not present.
Usually
 we assume that a forward wave will not 
 generate a significant backward wave via the nonlinearity 
 because the backward component is very poorly phase 
 matched\footnote{
 If the forward field has a wavevector $k_0$ 
 and evolves as $\exp(+\imath k_0 z)$, 
 the generated backward component will evolve as $\exp(-\imath k_0 z)$.
This gives a very rapid relative oscillation $\exp(-2\imath k_0 z)$, 
 which will quickly average to zero.
}. 
In contrast, 
 deliberately trying to phase match the backward wave was
 suggested in the 1960's \cite{Harris-1966apl}, 
 and some progress has been made in $\chi^{(2)}$ materials
 in recent years
 (e.g. 
 \cite{Ding-K-1996jqe,Kang-DBM-1997ol,Ding-KK-1998jqe,Sanborn-HD-2003josab}); 
 however such attempts are not the focus of this paper.

Here I use the phenomenon of carrier wave shocking 
 \cite{Rosen-1965pr} as a tool to probe 
 the fundamental limits of a uni-directional model
 under the influence of the intrinsically bi-directional nonlinear coupling.
Carrier wave shocks are discontinuities
 in the electric (and magnetic field) profiles, 
 and develop as the nonlinear effects distort the waveform as it propagates.
Assuming an initially uni-directional field, 
 I compare analytic predictions for the shocking distance
 from uni- and bi-directional theories.
These are based on the Method of Characteristics (MOC)
 \cite{Whitham-LWP}, 
 and show a clear difference between 
 the predicted shocking distances in the two cases.
The (exact) bi-directional theory is based on 
 the second order wave equation (as in 
 \cite{Rosen-1965pr,Kinsler-RTN-2007cshock}), 
 whereas
 the (approximate) uni-directional theory 
 is derived from the 
 $G^\pm$ directional fields \cite{Kinsler-RN-2005pra} description.
I support the theoretical results with 
 pseudospectral spatial domain (PSSD)
 simulations \cite{Tyrrell-KN-2005jmo}
 for both second order ($\chi^{(2)}$) and 
 third order ($\chi^{(3)}$) nonlinearities.

The results in this paper act as a bound on the validity of 
 one-dimensional 
 optical propagation models using a uni-directional approximation.
Since linear dispersion and finite nonlinear response times 
 will typically diminish any generation of a backward wave, 
 it is clear that any model which can be assumed uni-directional on 
 the basis of this paper will be more so in practise.
These results do not tell us whether
 the uni-directional approximation
 would be more or less robust for models incorporating transverse effects
 such as self-focussing, 
 but they at least establishes a point of reference,
 valid for beams with a weak spatial variation.
Note also that 
  I consider only propagation {\em within} bulk media, 
  since surfaces or interfaces can cause reflections, 
  and clearly require a bi-directional model.

Section \ref{S-standard} briefly describes the MOC, 
 and presents a prediction for the 
 shocking distance in the bi-directional case due to 
 simple nonlinearities of arbitrary order; 
 section \ref{S-directional} follows with analogous
 calculations using an explictly uni-directional 
 wave equation.
Next, 
 section \ref{S-comparisons} discusses the analytic 
 and numerical results,
 section \ref{S-backward} considers the role of the backward field, 
 and then section \ref{S-conclusions} presents some conclusions.

%
\section{Bi-directional model}
\label{S-standard}

Analytic formulae for the shocking distance in materials with 
 instantaneous response have been calculated for 
 both third-order nonlinearities (by Rosen \cite{Rosen-1965pr})
 and a simple second order case (by Radnor \cite{Radnor-2006}).
Both calculations used the MOC to 
 predict the formation of a value discontinuity in the field
 at certain points within the electric field profile of a pulse or wave.
Here I generalise the treatment to allow
 for an instantaneous perturbative nonlinearity of arbitrary order, 
 following the calculation of Kinsler et al. \cite{Kinsler-RTN-2007cshock}.
Note that the calculation below can easily be generalised to 
 include a sum of nonlinear terms, 
 if so desired \cite{Radnor-CKN-2007}.

\begin{figure}[ht]
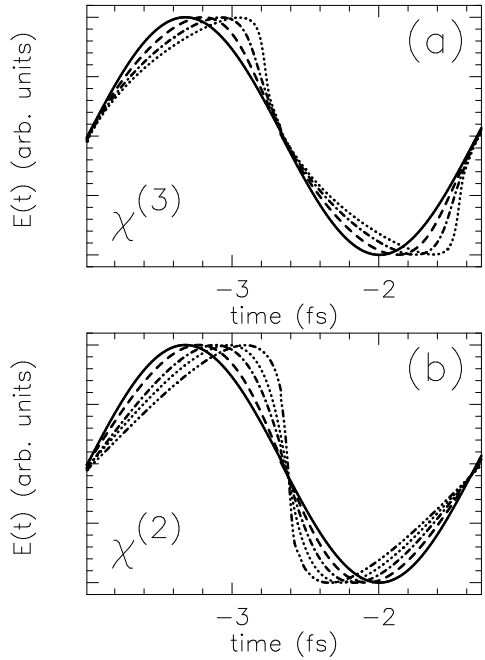

\includegraphics[width=0.50\columnwidth,angle=-90]{fig01a-chi3shock}
\includegraphics[width=0.50\columnwidth,angle=-90]{fig01b-chi2shock}
\caption{
The progressive 
 distortion of an initially sinusoidal wave profile as the 
 shocking distance is approached for
 (a) $\chi^{(3)}$ and (b) $\chi^{(2)}$ nonlinearities
 in a dispersionless medium.
In the spectral domain, 
 the distortion corresponds to a build-up of significant quantities 
 of higher order harmonic content.
}
\label{fig-chi3shock}
\end{figure}

In this paper,
 I consider nonlinearities of the simplest, 
 rather than the most general form, 
 because then solutions using the MOC can be found.
Ignoring the tensor nature of the coefficients, 
 the displacement field is
~
\begin{eqnarray}
 D 
&=&
  \epsilon_0 \left( E + \chi^{(1)} E + \chi^{(m)} E^m \right)
.
\label{eqn-Dexpansion}
\end{eqnarray}
The usual second order wave equation for $E$ 
 in this case is
~ 
\begin{eqnarray}
  c^2 \frac{\partial^2 E} {\partial z^2}
&=&
  n_0^2
  \frac{\partial^2 E }
       {\partial t^2}
 + \chi^{(m)} 
   \frac{\partial^2 E^m} 
        {\partial t^2}
,
\label{eqn-Ewaveeqn}
\end{eqnarray}
where $\epsilon_r = 1 + \chi^{(1)} = {n_0}^2$ is 
 the (relative) dielectric constant 
 and $n_0$ the linear refractive index.  

The MOC treats each point on the waveform $E(t)$ separately,
 and considers how it will move as the wave propagates.
The line traced out by the movement of any one of these points is a 
 characteristic.
The equation associated with eqn. (\ref{eqn-Ewaveeqn})
 that governs the characteristic lines of $E$ is
~
\begin{eqnarray}
  \frac{\partial E} 
       {\partial t} 
 + v_m ( E ) 
   \frac{\partial E} 
        {\partial z}
&=&
  0.
\label{eqn-characteristics}
\end{eqnarray}
with the velocity $v_m(E)$
  being that for a point on the wave with field strength $E$, 
  which is given by
~
\begin{eqnarray}
  v_m(E) 
&=&
  \frac {c }{ n_0} 
  \left[
    1 
   +
    m \chi^{(m)} E^{m-1} / n_0^2
  \right]^{-1/2}
.
\label {eqn-vE}
\end{eqnarray}
In this picture, 
  the nonlinearity manifests itself by giving different (fixed) velocities  
   to characteristics of different $E$.
  Thus, 
   as the wave profile $E(t)$ propagates forward in space ($z$), 
   the wave profile becomes distorted 
   by temporal compressions or expansions.
  A shock occurs if a region is compressed to the point where 
   two characteristics {\em intersect}.
  For a $\chi^{(3)}$ nonlinearity, 
   characteristics with higher $E^2$ move more slowly, 
   dragging the peaks toward later times; 
   a shock will first occur on the profile where $E^2$ is changing
   most rapidly in time.
  Waves approaching the point of shocking
   are shown in fig. \ref{fig-chi3shock}.

\begin{figure}[ht]
\includegraphics[width=0.50\columnwidth,angle=-0]{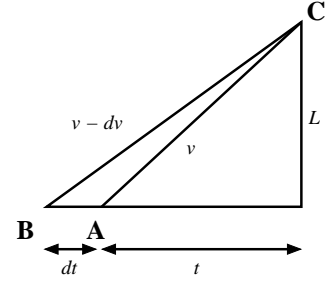}
\caption{
Here 
 two points A and B on the field profile (with fields $E_A, E_B$)
 follow their characteristics as the
 wave propagates.
Separated initially by a time difference $dt$, 
 they travel at different speeds ($v$ and $v-dv$), 
 and meet at point C.
}
\label{fig-moc}
\end{figure}

Using eqn. (\ref{eqn-vE}) along with 
 the construction shown in fig. \ref{fig-moc}, 
 we can derive a simple formula for 
 the distance to shocking. 
The figure shows two characteristics AC and BC, 
 originating from points A and B, 
 and converging towards a shock at C after a distance of $L$. 
 We have drawn the case where 
 the speed associated with AC (represented by its gradient)
 is lower than that of BC.  
From the geometry of the figure, 
 it is easy to show that
~
\begin{eqnarray}
  \frac{dv}{dt}
&=&
  \frac{v}{t}
=
  \frac{v^{2}}{L}
\label{eqn-moc1}
\end{eqnarray}
where $t$, 
 $v$, 
 and $L=vt$ are respectively time, 
 velocity, and distance.
Differentiating the velocity leads to
~
\begin{eqnarray}
  \frac{dv_m}{dt}
&=&
  -
  \frac{m c \chi^{(m)} / n_0^2} 
  {2 \left( 1 + m\chi^{(m)} E^{m-1} / n_0^2\right)^{3/2} }
  \frac{\partial \left(E^{m-1}\right)}{\partial t}
\\
&=&
 -
  \frac{m \chi^{(m)}}
       {2c^2}
  v_m^3
  \frac{\partial \left(E^{m-1}\right)}{\partial t}
,
\label{eqn-moc2}
\end{eqnarray}
and combining eqns. (\ref{eqn-moc1}) and (\ref{eqn-moc2})
 yields
~
\begin{eqnarray}
  L_m 
~=~
  v_m^2 / \frac{dv_m}{dt}
&=&
  \frac{2 c n_0 \sqrt{1 + m\chi^{(3)} E^{m-1}/n_0^2}}
       {m \chi^{(m)} (-\partial E^{m-1}/ \partial t)}
.
\label{eqn-moc3}
\end{eqnarray}
 Thus $L_m$ depends on $E(t)$, 
  and so will vary across the pulse profile.
 As the profile propagates and evolves, 
  so will the predictions for $L_m$,
  nevertheless, 
  the point at which shocking occurs 
  relative to the origin remains fixed.
For a given profile, 
 a shock will occur first at the point where the $L_m$
 reaches its minimum value.
We can therefore define the shocking distance, 
 for any arbitrary waveform $E(t)$, 
 as 
~
\begin{eqnarray}
  S_{m}
&=&
  \frac{2 c n_0}
       {m }
  \textrm{Min}
    \left[
      \frac{C_m / \chi^{(m)}}{ (-\partial E^{m-1}/ \partial t)}
    \right]
,
\label{eqn-moc4}
\\
\textrm{where}
~~~~
~~~~
  C_m
&=&
  \sqrt{1 + m \chi^{(m)} E^{m-1}/n_0^2}
.
\label{eqn-moc4-Cm}
\end{eqnarray}
We see that the shocking distance is inversely proportional 
 to the nonlinear strength $\chi^{(m)}$ (as would be expected), 
 and that the other important quantity
 is a derivative of powers of the field,
 i.e. ($\partial E^{m-1}/ \partial t$).
Note that for $\chi^{(m)}>0$,
 the nonlinear correction factor $C_m$ will 
 always be greater than $1$ for odd-order nonlinearities, 
 but may be less than $1$ for even-order nonlinearities.
In the limit of weak nonlinearity,
 $m \chi^{(m)} E^{m-1}/n_0^2 \ll 1$, 
~
\begin{eqnarray}
  S_{m}
&\simeq&
  \frac{2 c n_0}
       {m}
  \textrm{Min}
    \left[
      {\frac{1 / \chi^{(m)}}{(-\partial E^{m-1}/ \partial t)}}
    \right]
.
\label{eqn-moc4approx}
\end{eqnarray}

%
\section{Uni-directional model}
\label{S-directional}

Using the directional field approach of 
 Kinsler et al.\cite{Kinsler-RN-2005pra}
 we can write a pair of coupled 
 first order wave equations for directional $G^\pm$
 fields under the influence of an $m$-th order nonlinearity 
 without dispersion.
In the present case of a dispersionless medium, 
  these are defined by using 
  $G^\pm = \sqrt{\epsilon} E_x \pm \sqrt{\mu_0} H_y$, 
  where $\epsilon = \epsilon_0 ( 1 + \chi^{(1)} )$.
This combination of scaled fields provides us
  with a $G^+$ field whose Poynting vector points forward 
  along the $z$ axis, 
  and a $G^-$ field whose Poynting vector points backward. 
To get a uni-directional model, 
 we simply set the backward propagating $G^-$ field 
 to zero, 
 leaving us with a single first order wave equation
 for the forward propagating $G^+$ field.
Importantly, 
 we do not require the use of an exponential carrier function
 to impart the directionality 
 (see e.g. \cite{Casperson-1991pra,Sanborn-HD-2003josab}), 
 as this is achieved by the construction of the $G^\pm$ fields
 themselves.
As in section \ref{S-standard},  
 the calculation below can easily be generalised to 
 include a sum of nonlinear terms of different order, 
 if so desired.

After defining
 $E^\pm = G^\pm / 2\sqrt{\epsilon}$ \cite{Kinsler-2007envel}, 
 the coupled bi-directional wave equations are
~
\begin{eqnarray}
  \frac{\partial E_\pm}
       {\partial z}
 +
  \frac{n_0}{c}
  \frac{\partial E_\pm}
       {\partial t}
&=& 
 \pm
  \frac{\chi^{(m)}}{2 c} 
  \frac{\partial \left(E_+ + E_-\right)^m}
       {\partial t}
.
\label{eqn-fwd-waveeq}
\end{eqnarray}
Setting $E_-=0$ gives us a single, 
 uncoupled, 
 uni-directional wave equation
 which we can rewrite as
~
\begin{eqnarray}
  \frac{\partial E_+}
       {\partial t}
 +
  {v_{m+}(E_+)}
  \frac{\partial E_+}
       {\partial z}
&=&
 0
,
\label{eqn-fwd-only}
\end{eqnarray} 
with velocity
~
\begin{eqnarray}
  u_{m}(E_+) 
&=&
  \frac{c}{n_0}
  \left[
    1
   +
    m \chi^{(m)} E_+^{m-1} / 2 n_0^2
  \right]^{-1}
.
\label{eqn-fwd-vE}
\end{eqnarray}

Comparing eqn. (\ref{eqn-fwd-only}) to eqn. (\ref{eqn-characteristics})
 we see they have the same form;
 so that this wave equation {\em also} describes
 the motion of its characteristics.
Note also that eqn. (\ref{eqn-fwd-vE}) 
 is equivalent to a first-order expansion 
 of the square root term in eqn. (\ref{eqn-vE}).
We can again use the method of characteristics. 
Differentiating the velocity $u_{m}$ gives
~
\begin{eqnarray}
  \frac{du_{m}}{dt}
&=&
  -
  \frac{n_0}{c}
  \frac{m c \chi^{(m)} /  2 n_0^2 } 
  {\left( 1 + m\chi^{(m)} E_+^{m-1} / 2 n_0^2 \right)^{2} }
  \frac{\partial \left(E_+^{m-1}\right)}{ \partial t}
,
\label{eqn-fwd-moc2}
\end{eqnarray}
and combining eqns. (\ref{eqn-moc1}) and (\ref{eqn-fwd-moc2})
 yields the shocking distance
~
\begin{eqnarray}
  S_{m+}
&=&
  \frac{2 c n_0}
       {m }
  \textrm{Min}
    \left[
      \frac{1 / \chi^{(m)}}{ (-\partial E^{m-1}/ \partial t)}
    \right]
\label{eqn-fwd-moc4}
\end{eqnarray}
This is the same as the weak nonlinearity limit 
 of the bi-directional prediction in eqn. (\ref{eqn-moc4}), 
 i.e. it lacks the correction factor $C_m$
 defined in eqn. (\ref{eqn-moc4-Cm}).

It is important to note that eqn. (\ref{eqn-fwd-moc4})
 was found using an explicitly uni-directional formalism, 
 in which the {\em only} approximation was of 
 uni-directional propagation with no coupling to 
 the backward wave.
Thus it allows an unambiguous comparison of uni-directional 
 and bi-directional propagation.

%
\section{Discussion}
\label{S-comparisons}

We can see by comparing the shocking distances predicted 
 by the (exact) bi-directional theory in eqn. (\ref{eqn-moc4}) 
 and the (approximate) uni-directional theory in eqn. (\ref{eqn-fwd-moc4})
 that both theoretical predictions
 have the same two dominant trends:
 they are inversely dependent on the nonlinear coefficient $\chi^{(m)}$,
 and on the gradient of the $(m-1)$th power
 of the field.

The {\em difference} between the bi- and uni-directional theories 
 lies in the nonlinear correction factor $C_m$, 
 which only becomes significant for 
 extremely strong nonlinearities.
For small nonlinearities,
 and correspondingly long shocking distances, 
 the poorly phase matched backward component does not build up, 
 so the forward field (and hence shocking distances) are barely affected.
In fused silica near the damage threshold, 
 effective nonlinearities of order $\chi^{(3)} E^2 \simeq 0.06$
 can be achieved.
In the absence of dispersion,
 this amount of nonlinearity would lead to shocking distances of
 less than three wavelengths (e.g. $S_3 \sim 2\mu$m for $\lambda = 800$nm), 
 a regime in which $C_3 \simeq 1.04$.

%
%
%
%
%
%

For stronger nonlinearities, 
 significant conversion can take place before the field 
 has propagated even one wavelength, 
 let alone the several needed for the averaging effect 
 caused by poor phase matching.
The growth of a significant backward component then
 affects the propagation;
 indeed it can even be very strongly altered 
 if $m \chi^{(m)} E^{m-1}/n_0^2 \sim 1$.
However, 
 such extreme nonlinearities are not achievable
 in realistic systems, 
 since they require field intensities far beyond the damage thresholds 
 of standard nonlinear materials.
Nevertheless, 
 the point of this paper is to examine the limits
 of the uni-directional approximation: 
 adding more realistic material models would test only those models, 
 not the uni-directional approximation.

To support the theory,
 I have also done simulations of the bi-directional 
 and uni-directional cases using the PSSD technique \cite{Tyrrell-KN-2005jmo}.
These are (a) 
 straightforward simulations of Maxwell's equations, 
 which are naturally bi-directional;
(b)
 bi-directional simulations of Maxwell's equations, 
 using the $G^\pm$ directional fields \cite{Kinsler-RN-2005pra};
(c)
 uni-directional simulations, 
 using a $G^+$-only forward propagating model.
I treat nonlinear orders $m=2,3$ only 
 and propagate two-cycles of a (sinusoidal) CW field 
 sampled by 512 or 2048 points, 
 giving time resolutions of $dt=0.0128$fs and $0.0032$fs.
For the $E, H$ simulations I use a Yee-style \cite{Yee-1966tap}
 staggered grid and the PSSD method \cite{Tyrrell-KN-2005jmo}; 
 for the $G^\pm$ or $G^+$ simulations I use
 a leapfrog method, 
 which, 
 being analogous to the staggered grid, 
 ensures that the numerical performance of the simulations is 
 comparable.
Numerical ``shocks'' are detected by using the 
 local discontinuity detection (LDD)
 technique \cite{Kinsler-RTN-2007cshock}.

\begin{figure}
\includegraphics[height=0.80\columnwidth,angle=-90]{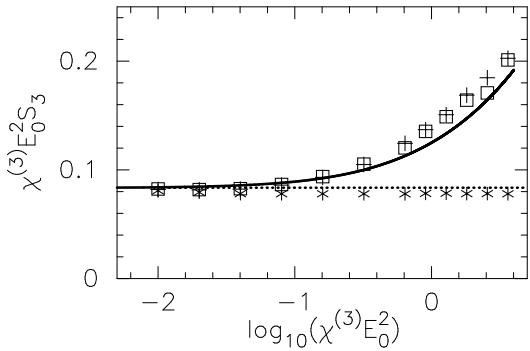}
\caption{
Scaled shocking distances for $\chi^{(3)}$
 as a function of nonlinearity:
 MOC predictions for the exact bi-directional case (solid line)
 uni-directional approximation (dotted line).
Simulation results are denoted by symbols:
 with $E,H$ ($\Box$), 
 $G^\pm$ ($+$), 
 uni-directional $G^+$ ($*$);  
 with distances determined using LDD shock detection.
}
\label{fig-pssd-Ls-chi3}
\end{figure}

The shocking distances 
 vary slightly from the comparable theory for two reasons.
First, 
 the LDD technique
 is not a perfect predictor of shocks.
Second,
 the staggered grids used by the simulations
 make it very hard
 to perfectly match the two initial fields values needed, 
 especially in the case of strong nonlinearity.

\begin{figure}
\includegraphics[height=0.80\columnwidth,angle=-90]{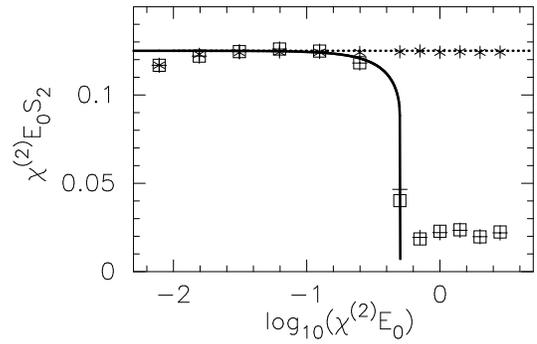}
\caption{
Scaled shocking distances for $\chi^{(2)}$
 as a function of nonlinearity:
 MOC predictions for the exact bi-directional case (solid line),
 uni-directional approximation (dotted line).
Simulation results are denoted by symbols, 
 with $E, H$ ($\Box$),
 $G^\pm$ ($+$), 
 uni-directional $G^+$ ($*$); 
 with distances determined using LDD shock detection.
}
\label{fig-pssd-Ls-chi2}
\end{figure}

Figs. \ref{fig-pssd-Ls-chi3} shows the behaviour of the 
 shocking distance 
 for the usual case of $\chi^{(3)}$ nonlinearity, 
 and fig. \ref{fig-pssd-Ls-chi2} shows it for the 
 alternative case of $\chi^{(2)}$ nonlinearity;
 both show 
 excellent agreement between the MOC theory and the simulations.
Note that the bi-directional simulations confirm the 
 bi-directional MOC theory,
 and the uni-directional simulations confirm the
 uni-directional MOC theory.
In particular, 
 the bi-directional simulations
 using $G^\pm$ fields agree remarkably well 
 with the Maxwell's equations simulations (as they should); 
 this serves to emphasise the difference between these
 and the contrasting uni-directional simulations
 based on (only) the $G^+$ field.

Note that these results
 are based on analytic results obtained using the MOC, 
 which cannot account for material dispersion.
Consequently, 
 while it is still possible to use them to validate (or not) the
 use of a uni-directional approximation in a physical model, 
 as I do here, 
 it is hard to see how they could be experimentally tested.
Any experiment would instead need to compare its results with simulations
 of the chosen nonlinear material 
 that included linear dispersion, 
 an accurate model of the nonlinear response, 
 and perhaps even transverse effects.
The test would then be a match with a bi-directional simulation, 
 but a disagreement with a uni-directional simulation.

%
\section{The role of the backward field}
\label{S-backward}

Figs. \ref{fig-pssd-Ls-chi3} and \ref{fig-pssd-Ls-chi2} exhibit
 different trends for the scaled shocking distances
 $\chi^{(3)} S_{3}$ and $\chi^{(2)} S_{2}$; 
 a difference due to the nonlinear correction $C_m$, 
 which contains a field dependent part $F_m = m \chi^{(m)} E^{m-1}/n_0^2$.
In the even-order case
 $F_m$ can take either sign;
 whereas for odd-order it can only have the sign of $\chi^{(m)}$.
Thus for sufficiently strong nonlinearities,
 $C_m$ could become zero, 
 implying a {\em zero} shocking distance.
This can happen for odd-order nonlinearities
 only if $\chi^{(m)}$ is negative, 
 but will always be possible in the even-order case.
This exotic ``instant shock'' behaviour is a mathematical 
 prediction rather than a physical one, 
 since the effective nonlinearities required are far in excess of
 those attainable in experiment, 
 and realistic nonlinearities
 do not have an instantaneous response.
Nevertheless, 
 it is instructive to examine
 the reason for this surprising effect, 
 as it demonstrates how backward waves can influence the forward ones.

We see this instant shock regime in 
 fig. \ref{fig-pssd-Ls-chi2}, 
 where the trend for $\chi^{(2)} S_{2}$ is downward, 
 with a sudden dip toward zero when the peak field $E_0$ gives $F_2=-1$, 
 and remaining at zero for still stronger nonlinearities.
This causes numerical difficulties, 
 leading to relatively poor agreement between theory and simulation
 in this region 
 on fig. \ref{fig-pssd-Ls-chi2}; 
 although this improves 
 as the temporal resolution is increased.

\begin{figure}
\includegraphics[height=0.65\columnwidth,angle=-90]{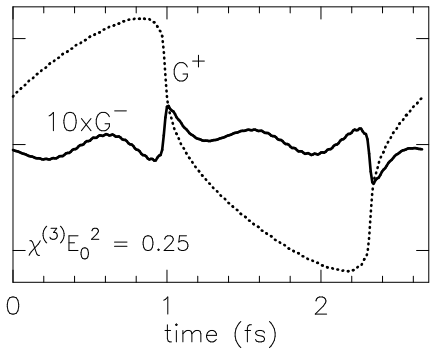}
\caption{
Comparison of forward ($G^+$) and backward ($G^-$) field contributions
 at the LDD shocking point 
 for a $\chi^{(3)}$ nonlinearity with $C_3 \simeq 1.32$.
The $G^-$ field has been scaled up
 to enhance detail.
}
\label{fig-pssd-Gpm-chi3}
\end{figure}

On fig. \ref{fig-pssd-Gpm-chi3} we see the
 forward and backward field components $G^\pm$ 
 in the case of a positive $\chi^{(3)}$, 
 where $\chi^{(3)} S_{3}$ increases with nonlinearity.
Both are steepest near $\sim 1$fs, 
 where
 $G^-$ has a gradient of the opposite sign to that of $G^+$, 
 so the gradient of $E \sim G^+ + G^-$ is reduced.
Consequently the backward $G^-$ wave has the effect
 of {\em increasing} the shocking distance; 
 although the effect would reverse
 for a negative $\chi^{(3)}$.
Note that both these components are strongly coupled to each other.

\begin{figure}
\includegraphics[height=0.65\columnwidth,angle=-90]{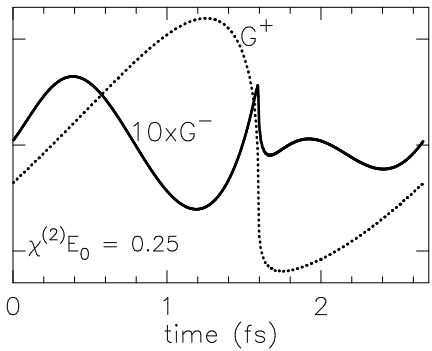}
\caption{
Comparison of forward ($G^+$) and backward ($G^-$) field contributions
 at the LDD shocking point 
 for a $\chi^{(2)}$ nonlinearity with $C_2 \simeq 0.71$.
The $G^-$ field has been scaled up
 to enhance detail.
}
\label{fig-pssd-Gpm-chi2}
\end{figure}

In contrast, 
 in the $\chi^{(2)}$ case
 on fig. \ref{fig-pssd-Gpm-chi2},
 the equivalent region (near $1.3$fs) shows that the 
 gradients of $G^+$ and $G^-$ can have the same sign, 
 enhancing the gradient of $E$.
Closer inspection shows that the gradient of $G^-$ abruptly 
 switches sign where $G^+$ has a point of inflection, 
 so the $G^-$ field mitigates shocking on one side, 
 and enhances it on the other.
The enhancement then acts like a feedback process, 
 where the backward wave steepens the gradient, 
 which in turn enhances the backward wave, 
 and so on.
Thus not only does the backward wave {\em decrease}
 the shocking distance, 
 but for sufficiently strong nonlinearities the 
 feedback can cause a shock in 
 an infinitesmally short distance -- 
 just as predicted by eqn. (\ref{eqn-fwd-moc4}).

Note that
 since the effect of the $\chi^{(2)}$ nonlinearity depends
 on the sign of the field, 
 the backward field components will tend to increase the 
 shocking distances in some regions of the field profile, 
 but reduce them in others. 
However, 
 since we always look for the minumum shocking distance,
 this will always be reduced,
 whatever the sign of $\chi^{(2)}$.

%
\section{Conclusions}
\label{S-conclusions}

I have demonstrated the fundamental limits of the 
 widely used uni-directional propagation approximation.
This was done by comparing analytic results 
 for the shocking distance obtained from
 both an exact bi-directional model,
 and an approximate uni-directional model.
These theoretical results
 were for simple nonlinearities of arbitrary order, 
 based on the method of characteristics, 
 and are 
 supported by numerical simulations of both models.
The exact bi-directional results were based on 
 Maxwell's equations using $E$ \& $H$.
The approximate uni-directional results relied on the 
 construction of the $G^\pm$ directional fields \cite{Kinsler-RN-2005pra}, 
 which enabled the forward-only approximation
 to be made without introducing {\em any} additional assumptions.

The results show that the condition
 $| m \chi^{(m)} E^{m-1}/n_0^2 | \ll 1$
 must hold for the uni-directional approximation to be true; 
 even when no backward field is initially present.
This condition is usually easily satisified
 in nonlinear optical materials -- 
 even with fields strong enough to approach the damage threshold.
If this condition does not hold,
 then significant {\em non phase-matched} 
 forward-backward coupling can occur, 
 affecting the propagation accordingly; 
 in extreme cases causing behaviour like the ``instant shock''
 discussed in section \ref{S-backward}.
Such features, 
 demonstrated in a comparison between two models which {\em only}
 differ by a uni-directional approximation made in one, 
 provide an important indication of the limitations of a 
 uni-directional description.

%
\section*{Acknowledgments}

I acknowledge useful discussions 
 with G.H.C. New, S.B.P. Radnor, and J.M. Dudley.

~

~

%

\begin{widetext}

\end{widetext}

%
\section*{Appendix: Time propagated MOC}

%
\subsubsection{Bi-directional form}

Here we take the expression in eqn. (\ref{eqn-characteristics}), 
 and swap the roles of $t$ and $z$.
The associated equation 
 governing the characteristic lines of $E(z)$ is
~
\begin{eqnarray}
   w_m ( E ) 
  \frac{\partial E} 
       {\partial t} 
 + 
   \frac{\partial E} 
        {\partial z}
&=&
  0.
\label{eqn-t-characteristics}
\end{eqnarray}
with the (inverse) velocity $w_m(E)$ given by
~
\begin{eqnarray}
  w_m(E) 
&=&
  \frac { n_0} {c }
  \left[
    1 
   +
    m \chi^{(m)} E^{m-1} / n_0^2
  \right]^{1/2}
.
\label {eqn-t-vE}
\end{eqnarray}

Using eqn. (\ref{eqn-t-vE}) along with 
 as time-space swapped version of the 
  construction shown in fig. \ref{fig-moc}, 
 we can derive a simple formula for 
 the time to shocking. 
From the geometry, 
 it is easy to show that
~
\begin{eqnarray}
  \frac{dw}{dz}
&=&
  \frac{w}{z}
=
  \frac{w^{2}}{T}
\label{eqn-t-moc1}
\end{eqnarray}
where $z$, 
 $w$, 
 and $T=wz$ are respectively distance, 
 inverse velocity, and time.
Differentiating the inverse velocity leads to
~
\begin{eqnarray}
  \frac{dw_m}{dz}
&=&
  \frac { n_0} {2 c }
  \frac{ m \chi^{(m)}}{ n_0^2 }
  \left( 1 + m\chi^{(m)} E^{m-1} / n_0^2\right)^{-1/2} 
  \frac{dE^{m-1}}{dz} 
~~~~
\\
&=&
  \frac { 1 }{2}
  \frac { n_0^2} {c^2 }
  \frac{ m \chi^{(m)}}{ n_0^2 }
  w_m^{-1}
  \frac{dE^{m-1}}{dz}
\\
&=&
  \frac{ m \chi^{(m)}}{ 2 c^2}
  w_m^{-1}
  \frac{dE^{m-1}}{dz} 
,
\label{eqn-t-moc2}
\end{eqnarray}
and combining eqns. (\ref{eqn-t-moc1}) and (\ref{eqn-t-moc2})
 yields
~
\begin{eqnarray}
  T_m 
&=&
  w_m^2 / \frac{dw_m}{dz}
\\
&=&
  w_m^3
  \frac{ 2 c^2}{ m \chi^{(m)}}
  \left(
    \frac{dE^{m-1}}{dz} 
  \right)^{-1}
.
\label{eqn-t-moc3}
\end{eqnarray}
For a given profile, 
 a shock will occur first at the point where $T$
 reaches its minimum value.
We can therefore define the shocking time $\mathscr{T}_m$, 
 for any arbitrary waveform $E(z)$, 
 as 
~
\begin{eqnarray}
  \mathscr{T}_m
&=&
  \frac{2 n_0^3}
       {m c }
  Min
    \frac{C_m^3 / \chi^{(m)}}{ dE^{m-1}/dz}
,
\label{eqn-t-moc4}
\\
\textrm{where}
~~~~
~~~~
  C_m
&=&
  \sqrt{1 + m \chi^{(m)} E^{m-1}/n_0^2}
\label{eqn-t-moc4-Cm}
\\
  C_m^3
&\simeq&
  1 + \frac{3}{2} m \chi^{(m)} E^{m-1}/ n_0^2
.
\end{eqnarray}

In the spatially propagated form (eqn.(\ref{eqn-moc4})), 
 the correction term is $C_m$, 
 not the $C_m^3$ seen here.

Note that the size of the first order correction term is
 {\em independent} of the nonlinear strength, 
 and depends only on the properties of the field profile:
~
\begin{eqnarray}
  \delta \mathscr{T}_m
&=&
  \frac{2 n_0^3}
       {m c \chi^{(m)}}
 \times
  \frac{3 m \chi^{(m)} E^{m-1}}{2 n_0^2}
  \left(
    \frac{dE^{m-1}}{dz} 
  \right)^{-1}
\\
&=&
  \frac{6 n_0^3 m \chi^{(m)} E^{m-1} }
       {2 n_0^2 m c \chi^{(m)}}
 \times
  \left(
    \frac{dE^{m-1}}{dz} 
  \right)^{-1}
\\
&=&
  \frac{3 n_0  E^{m-1} }
       { c }
 \times
  \left(
    \frac{dE^{m-1}}{dz} 
  \right)^{-1}
.
\end{eqnarray}

%
\subsubsection{Uni-directional form}

Here we take the expression in eqn. (\ref{eqn-characteristics}), 
 and swap the roles of $t$ and $z$.
The associated equation 
 governing the characteristic lines of $E(z)$ is
~
\begin{eqnarray}
   w_{m+} ( E ) 
  \frac{\partial E} 
       {\partial t} 
 + 
   \frac{\partial E} 
        {\partial z}
&=&
  0.
\label{eqn-t-fwd-characteristics}
\end{eqnarray}
with the (inverse) velocity $w_{m+}(E)$ given by
~
\begin{eqnarray}
  w_{m+}(E) 
&=&
  \frac { n_0} {c }
  \left[
    1 
   +
    m \chi^{(m)} E^{m-1} / 2 n_0^2
  \right]
.
\label {eqn-t-fwd-vE}
\end{eqnarray}

Using eqn. (\ref{eqn-t-vE}) along with 
 as time-space swapped version of the 
  construction shown in fig. \ref{fig-moc}, 
 we can derive a simple formula for 
 the time to shocking. 
From the geometry, 
 it is easy to show that
~
\begin{eqnarray}
  \frac{dw}{dz}
&=&
  \frac{w}{z}
=
  \frac{w^{2}}{T}
\label{eqn-t-fwd-moc1}
\end{eqnarray}
where $z$, 
 $w$, 
 and $T = w z$ are respectively distance, 
 inverse velocity, and time.
Differentiating the inverse velocity leads to
~
\begin{eqnarray}
  \frac{dw_{m+}}{dz}
&=&
  \frac { n_0} {c }
  \frac{ m \chi^{(m)}}{ 2 n_0^2 }
  \left( 1 + m\chi^{(m)} E^{m-1} / 2 n_0^2\right)^0
  \frac{dE^{m-1}}{dz} 
~~~~
\\
&=&
  \frac { n_0} {c }
  \frac{ m \chi^{(m)}}{ 2 n_0^2 }
  \frac{dE^{m-1}}{dz}
\\
&=&
  \frac{ m \chi^{(m)}}{ 2 n_0 c^2}
  \frac{dE^{m-1}}{dz} 
,
\label{eqn-t-fwd-moc2}
\end{eqnarray}
and combining eqns. (\ref{eqn-t-moc1}) and (\ref{eqn-t-moc2})
 yields
~
\begin{eqnarray}
  T_{m+}
&=&
  w_{m+}^2 / \frac{dw_{m+}}{dz}
\\
&=&
  w_{m+}^2
  \frac{ 2 n_0 c}{ m \chi^{(m)}}
  \left(
    \frac{dE^{m-1}}{dz} 
  \right)^{-1}
\\
&=&
  \frac{n_0^2}{c^2}
  \left( 1 + m\chi^{(m)} E^{m-1} / 2 n_0^2\right)^2
  \frac{ 2 n_0 c}{ m \chi^{(m)}}
  \left(
    \frac{dE^{m-1}}{dz} 
  \right)^{-1}
.
\label{eqn-t-fwd-moc3}
\end{eqnarray}
For a given profile, 
 a shock will occur first at the point where $T$
 reaches its minimum value.
We can therefore define the shocking time $\mathscr{T}_{m+}$, 
 for any arbitrary waveform $E(z)$, 
 as 
~
\begin{eqnarray}
  \mathscr{T}_{m+}
&=&
  \frac{2 n_0^3}
       {m c }
  Min
    \frac{D_m^2 / \chi^{(m)}}{ dE^{m-1}/dz}
,
\label{eqn-t-fwd-moc4}
\\
\textrm{where}
~~~~
~~~~
  D_m
&=&
  1 + m \chi^{(m)} E^{m-1}/2 n_0^2
\label{eqn-t-fwd-moc4-Cm}
\\
  D_m^2
&\simeq&
  1 + m \chi^{(m)} E^{m-1}/ n_0^2
.
\end{eqnarray}

In the spatially propagated form (eqn.(\ref{eqn-fwd-moc4})), 
 the $D_m$-like term is simply $1$.

Note that the size of the first order correction term is
 {\em independent} of the nonlinear strength, 
 and depends only on the properties of the field profile:
~
\begin{eqnarray}
  \delta \mathscr{T}_{m+}
&=&
  \frac{2 n_0^3}
       {m c \chi^{(m)}}
 \times
  \frac{ m \chi^{(m)} E^{m-1}}{ n_0^2}
  \left(
    \frac{dE^{m-1}}{dz} 
  \right)^{-1}
\\
&=&
  \frac{2 n_0^3 m \chi^{(m)} E^{m-1} }
       { n_0^2 m c \chi^{(m)}}
 \times
  \left(
    \frac{dE^{m-1}}{dz} 
  \right)^{-1}
\\
&=&
  \frac{2 n_0  E^{m-1} }
       { c }
 \times
  \left(
    \frac{dE^{m-1}}{dz} 
  \right)^{-1}
.
\end{eqnarray}

%
\subsubsection{Comparison: Bi- vs Uni-directional}

The comparison is not as neat as for the spatially propagated case, 
 but nevertheless is very similar.
For the small-nonlinearity limit (i.e. $m \chi^{(m)} E^{m-1}/ n_0^2 \ll 1$), 
 the difference is just the factor of $1 + m \chi^{(m)} E^{m-1}/2 n_0^2$
 that appears for the spatially propagated results in the same limit.

Note that the shocking time $\mathscr{T}$ is not just related to the 
 shocking distance $L$ by a simple factor of $c$, 
 since in the spatially propagated picure, 
 the peak-dragging effect corresponds to a time offset which needs 
 to be added to the propagation time $L/c$.

\end{document}